\documentclass[reprint, NumberedRefs]{JASA}

\begin{document}
%% the square bracket argument will send term to running head in
%% preprint, or running foot in reprint style.

\title[]{}

\title[Music source separation]{Music source separation conditioned on 3D point clouds}
\affiliation{Department of Music Acoustics, University of Music and Performing Arts Vienna, Austria}
\author{Francesc Llu\'is}
\email{lluis-salvado@mdw.ac.at}
\author{Vasileios Chatziioannou}
\author{Alex Hofmann}

%\preprint{Author, JASA}		%  if you want want this message to appear in upper left corner of title page

\date{\today}

\begin{abstract}
Recently, significant progress has been made in audio source separation by the application of deep learning techniques. 
Current methods that combine both audio and visual information use 2D representations such as images to guide the separation process.
However, in order to (re)-create acoustically correct scenes for 3D virtual/augmented reality applications from recordings of real music ensembles, detailed information about each sound source in the 3D environment is required.
This demand, together with the proliferation of 3D visual acquisition systems like LiDAR or rgb-depth cameras, stimulates the creation of models that can guide the audio separation using 3D visual information.
This paper proposes a multi-modal deep learning model to perform music source separation conditioned on 3D point clouds of music performance recordings. This model extracts visual features using 3D sparse convolutions, while audio features are extracted using dense convolutions. A fusion module combines the extracted features to finally perform the audio source separation.
It is shown, that the presented model can distinguish the musical instruments from a single 3D point cloud frame, and perform source separation qualitatively similar to a reference case, where manually assigned instrument labels are provided.

\end{abstract}

%% pacs numbers not used

\maketitle

\maketitle

\section{\label{sec:1} Introduction}

The task of recovering each audio signal contribution from an acoustic mixture of music has been actively explored throughout the years\cite{cano2018musical} and significant progress has been made recently using deep learning methods\cite{defossez2019music, stoter2019, samuel2020meta, lluis2019end}. Although music source separation has been traditionally tackled using only audio signals, the emergence of data-driven techniques has facilitated the inclusion of visual information to guide models in the audio separation task.
Music source separation conditioned on visual information has also gained significant improvement through the application of deep learning models\cite{zhu2020deep}.
The proposed deep learning methods mainly use appearance cues from the 2D visual representations to condition the separation. Gao et al.\cite{gao2018learning} proposed a two step procedure, where initially a neural network learns to associate audio frequency bases with musical instruments appearing in a video. Then, the learned audio bases guide a non-negative matrix factorization framework to perform source separation. Concurrently, Zhao et al.\cite{zhao2018sound} proposed PixelPlayer, a method capable to jointly learn from audio and video to perform both source separation and localization. Since then, source separation models conditioned on visual information have been enhanced with different approaches such as using recursivity\cite{xu2019recursive} or co-separation loss\cite{gao2019co}. Recently the importance of appearance cues for music source separation\cite{zhu2020separating} and multi-modal fusion strategies\cite{slizovskaia2020conditioned} have been studied. In addition, other methods learning also from motion representation, such as low-level optical flow \cite{zhao2019sound} or keypoint-based structured representation\cite{gan2020music}, have been proposed to facilitate the model to separate simultaneously musical sources coming from the same type of instrument.

The proliferation of 3D visual acquisition systems such as LiDAR and various rgb-depth cameras in devices used in a day-to-day fashion allows to easily capture 3D visual information of the environment. This, together with their capability to record audio, stimulates the development of algorithms that can learn from 3D video along with audio and opens new possibilities for virtual reality (VR) and augmented reality (AR) applications. Audio is crucial to bring meaningful experiences to people in VR/AR applications. VR/AR applications are meant to be a multimodal and interactive experience where the sound perception has to be convincing and match the other sensory information perceived. A first step to render proper acoustic stimuli for VR/AR applications is to know the sound of each object in the 3D environment\cite{vorlander2008auralization}. When scanning and recording an environment, multiple sources emit sound simultaneously and therefore, source separation algorithms are needed to separate the audio and associate it to each source in the 3D environment.

In this paper, we propose a model for source separation conditioned on 3D videos, i.e.\ sequences of 3D point clouds, an approach which appears to be unexplored. For clarity of exposition, the deep learning model learns to separate individual sources from a musical audio mixture given the 3D video associated with the source of interest. The model consists of a 3D video analysis network which uses sparse convolutions to extract visual features from the 3D visual information and an audio network with dense convolutions that extracts audio spectral features from an audio mixture spectrogram. Finally, an end module leverages the multimodal extracted features to perform source separation. 3D visual representations, as opposed to 2D representations, allow to access the location of the sources in space and therefore exploit the distance between each source and the receiver for further auralization purposes\cite{vorlander2008auralization}. However, unlike 2D visual representations which are represented in a dense grid, 3D representations are very sparse and the surface of the captured objects are represented as a set of points in the 3D space which adds the additional difficulty when extracting local and global geometric patterns.

This paper is organized as follows: Section \ref{sec:Approach} contains a detailed explanation of each part of the model as well as the data used and the data processing applied. Section \ref{sec:Results} presents the implementation details, the evaluation metrics, and the experimental results. Finally, Section \ref{sec:Discussion} provides a discussion on the obtained results.

\section{\label{sec:Approach} Approach}

We propose a learning algorithm capable of separating an individual sound source contribution from an audio mixture signal given the 3D video associated with it. Considering an audio mixture signal
\begin{equation}
s_\mathrm{mix}(t)=\sum_{i=1}^{N}s_i(t)
\end{equation}
generated by $N$ sources $s_i$ and its associated 3D video $v_i$, we aim at finding a model $f$ with the structure of a neural network such that $s_i(t)=f(s_\mathrm{mix}(t), v_i)$.  
The proposed model takes as inputs the magnitude spectrogram of the audio mixture along with the 3D video of the source we aim to separate and predicts a spectrogram mask. The separated waveform $\hat{s}_i$ is obtained after performing the inverse short-time Fourier transform (iSTFT) to the multiplication of the predicted spectrogram mask and the input mixture spectrogram. Note that masked-based separation is commonly used in source separation algorithms operating in the time-frequency domain as it works better than direct prediction of the magnitude spectrogram\cite{wang2014training}.

\begin{figure*}[!t]
	\centering
	\includegraphics[width=\linewidth]{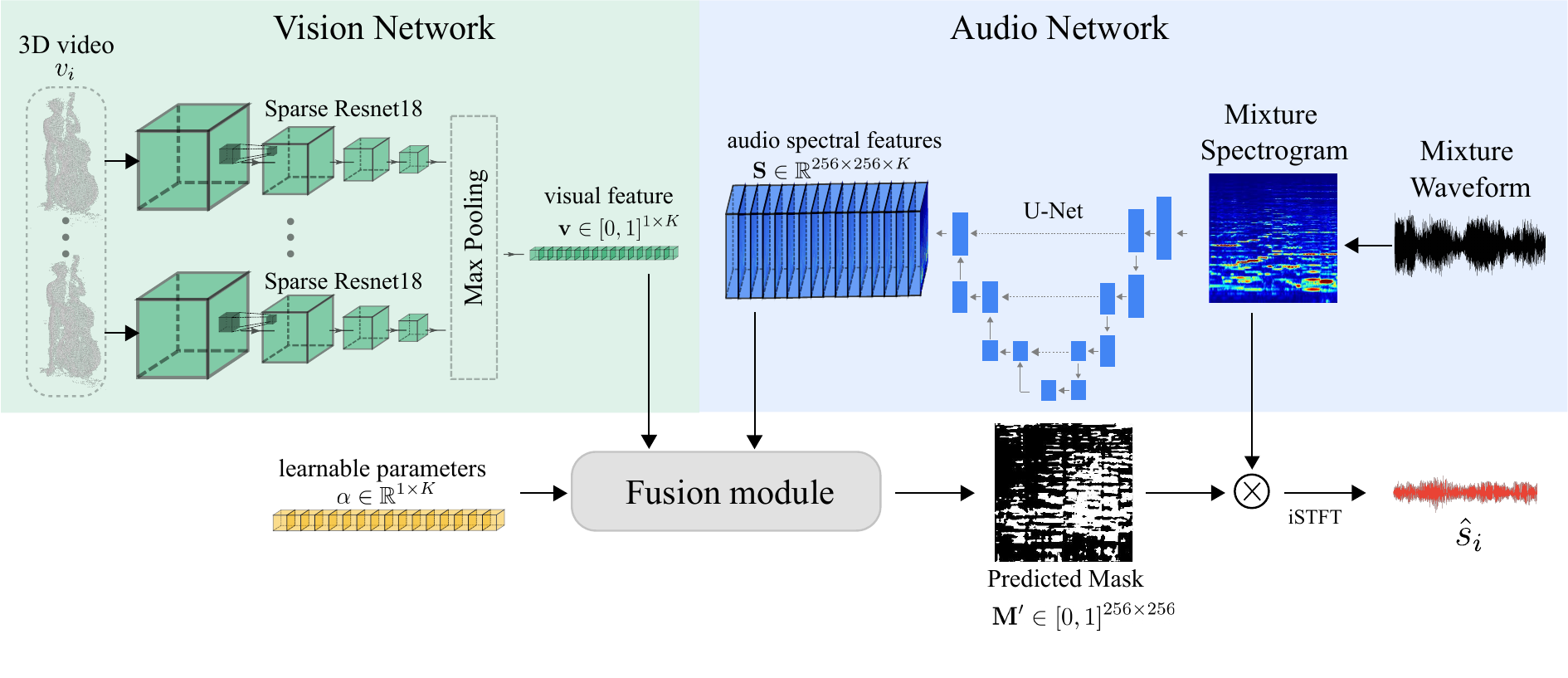}
	\caption{Overview diagram, showing the different parts of the model. Green area represents the vision network which extracts a visual feature after analyzing each 3D video frames using a sparse Resnet18 architecture. Blue area represents the audio network which extracts audio spectral features after analyzing the audio mixture spectrogram using a U-Net architecture. The fusion module combines the multimodal extracted features to perform source separation.
	}
	\label{fig:main_diagram}
\end{figure*}

The neural network architecture is adapted from the PixelPlayer\cite{zhao2018sound} and it consists of a 3D vision network, an audio network, and a fusion module. Broadly, the audio network extracts audio spectral features from the input spectrogram using 2D dense convolutions while the 3D vision network uses sparse 3D convolutions\cite{BMVC2015_150} to extract a visual feature vector from the 3D video of the source we aim to separate. Then, the fusion module combines the multimodal extracted features to predict the spectrogram mask. Fig.~\ref{fig:main_diagram} shows a schematic diagram of the proposed model.

For this research, we adopt the \textit{mix-and-separate}\cite{zhao2018sound, gan2020music, slizovskaia2020conditioned} learning approach. In a first step, the \textit{mix-and-separate} approach consists of generating artificial audio mixtures via mixing individual sound sources. Then, the learning objective is to recover each individual sound source conditioned on its associated visual information. This allows to use unlabeled individual sounds itself as supervision. Therefore, although the network is trained in a supervised fashion, the whole pipeline is considered self-supervised.

Music source separation conditioned on 3D visual information is a new domain. While there are lots of audio datasets\cite{fonseca2017freesound}, there is data scarcity of 3D videos from musicians. For the purposes of this work, we capture 3D videos of several performers playing five different instruments: cello, doublebass, guitar, saxophone, and violin. In addition, we separately collect audio recordings for these instrument types from existing audio datasets. Lastly, once all 3D video and audio data is collected, we randomly associate audio and 3D video for each instrument.

The data and the code of the proposed model is available online (see supplemental material) \footnote{See supplementary material at \url{github.com/francesclluis/point-cloud-source-separation}}.

\subsection{\label{subsec:3drepresentation}3D video representation with sparse tensors}

3D videos are a continuous sequence of 3D scans captured by rgb-depth cameras or LiDAR scanners. For each frame of the video, a point cloud can be used to represent the captured data.

A point cloud is a low-level representation that simply consist of a collection of points $\{P_i\}_{i=1}^I$. The essential information associated to each point is its location in space. For example, in a Cartesian coordinate system each point $P_i$ is associated with a triple of coordinates $\mathbf{c}_i\in\mathbb{R}^{3}$ in the x,y,z-axis. In addition, each point can have associated features $\mathbf{f}_i\in\mathbb{R}^{n}$ like its color.

A third order tensor $\mathbf{T}\in\mathbb{R}^{N_1\times N_2\times N_3}$ can be used to represent a point cloud. To this end, point cloud coordinates are discretized using a voxel size $v_s$ which defines coordinates in a integer grid, i.e.\ $\mathbf{c}_i^\prime = \lfloor\frac{\mathbf{c}_i}{v_s}\rfloor$. Given that point clouds contain empty space, this results in a sparse tensor where a lot of elements are zero.
\begin{equation}
\mathbf{T}[\mathbf{c}_i^\prime]=
\left\{
	\begin{array}{ll}
		\mathbf{f}_i  & \mbox{if } \mathbf{c}_i^\prime\in\textit{C}\\
		0 & \mbox{otherwise},
	\end{array}
\right.
\end{equation}
where $C=\textrm{supp}(\mathbf{T})$ is the set of non-zero discretized coordinates and $\mathbf{f}_i$ is the feature associated to $P_i$ with discretized coordinates $\mathbf{c}_i^\prime$.
In this work, 3D videos are represented as a sequence of sparse tensors where each tensor contains the frame's 3D information in the form of a point cloud.

\subsection{\label{subsec:visionnetwork}Vision Network}

The vision network consists of a Resnet18\cite{he2016deep} architecture with sparse 3D convolutions\cite{BMVC2015_150} to extract visual features from 3D video frames. We first introduce the sparse 3D convolution operation and subsequently present the main characteristics of the sparse Resnet18 architecture.

\subsubsection{Sparse 3D convolution}

The crucial operation within the vision network is the sparse 3D convolution. Considering predefined input coordinates $C_\mathrm{in}$ and output coordinates $C_\mathrm{out}$, convolution on a 3D sparse tensor can be defined as:
\begin{equation}
\mathbf{T}^\prime[x, y, z]= \sum_{i,j,k\in\mathcal{N}(x,y,z)} \mathbf{W}[i,j,k]\mathbf{T}[x+i,y+j,z+k]
\end{equation}
for $(x,y,z)\in C_\mathrm{out}$. Where  $\mathcal{N}(x,y,z)=\{(i,j,k)||i|\leq L, |j|\leq L, |k|\leq L, (i+x,j+y,k+z)\in C_{in}\}$. $\mathbf{W}$ are the weights of the 3D kernel and $2L+1$ is the convolution kernel size. Sparse 3D convolutions provide the necessary flexibility to efficiently learn from non-dense representations of variable size such as point clouds. Unlike images or 2D videos where the information is densely represented in a grid, point clouds contain lots of empty space. For this reason, it is inefficient to operate on them using traditional convolutions.

\subsubsection{Vision architecture}
\begin{figure}[!t]
	\centering
	\figline{
	\leftfig{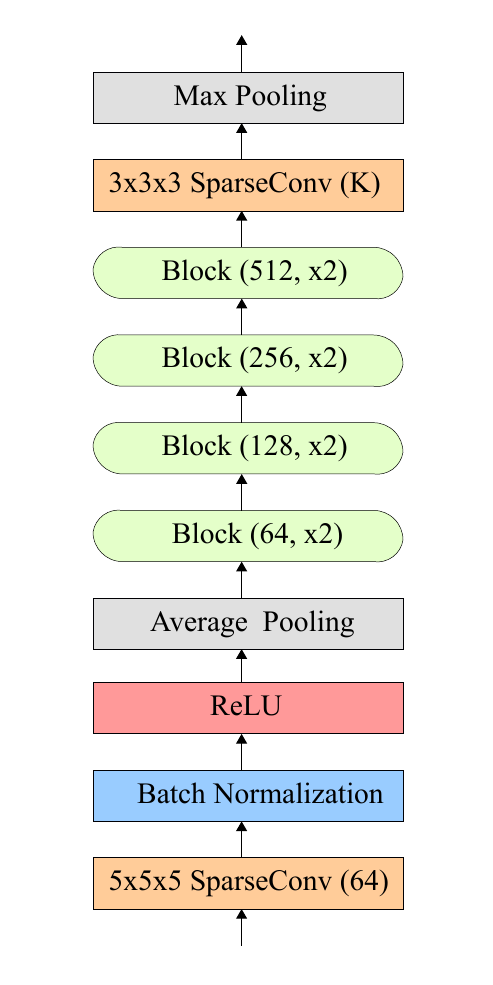}{0.5\linewidth}{(a)}
	\rightfig{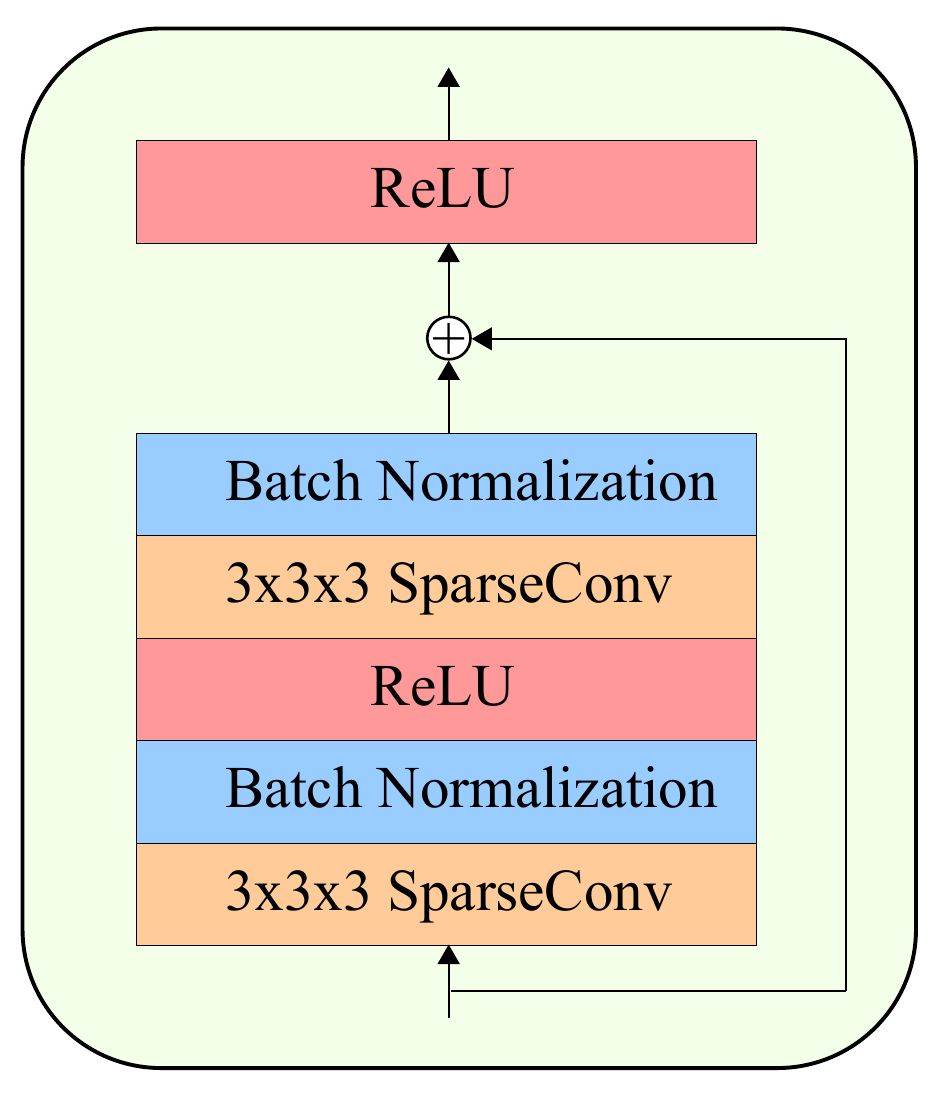}{0.5\linewidth}{(b)}}
	\caption{(a) Schematic diagram of the sparse Resnet18 architecture used to extract features for each 3D video frames. (b) Schematic diagram of the sparse Resnet18 residual block.}
	\label{fig:vision_network}
\end{figure}

We use a Resnet18 neural network \cite{he2016deep} with sparse 3D convolutions \cite{BMVC2015_150} to extract features from each frame of a 3D video. Resnet18 architecture was first introduced for the task of image recognition\cite{he2016deep} and since then it has been successfully used in several computer vision tasks\cite{khan2020survey}.
Key to the Resnet18 architecture are the skip connections within its residual blocks which assist convergence with negligible computational cost. The skip connection at the beginning of a residual block is then added at the end which helps to propagate low-level information through the network. To learn at different scales, the feature space is halved after two residual blocks and the receptive field is doubled by using a stride of 2. In addition, the depth of the filter size is doubled. Through the network, ReLU is used as activation function and batch normalization is applied after sparse convolutions. At the top of the network we add a 3x3x3 sparse convolution with $K$ output channels and apply a max-pooling operation to adequate the dimensions of the extracted features. Figure~\ref{fig:vision_network} shows a schematic diagram of the Resnet18 architecture with sparse 3D convolutions.

Finally, another max-pooling operation with sigmoid activation is applied to all frame features which results in an extracted visual feature vector $\mathbf{v}\in[0,1]^{1\times K}$ (See Fig.~\ref{fig:main_diagram}).

\subsection{\label{subsec:audionetwork}Audio Network}

We use a U-Net architecture with dense convolutions to extract spectral features from the audio mixture spectrogram. U-Net was first introduced for biomedical image segmentation\cite{ronneberger2015u} and since then it has been adapted for audio source separation in many cases\cite{jansson2017singing, zhao2018sound, gao2019co}. The U-Net encoder-decoder structure learns multi-resolution features from the magnitude spectrogram. The encoder halves the size of the feature maps by using a stride of 2 and doubles the filter size while the decoder then reverses this procedure by upsampling the size of the feature maps and halving the filter size. Similarly to Resnet18, skip connections are also present in U-Net which allows low-level information to be propagated through the network. The U-Net feature maps computed by the encoder are accessed by the decoder at the same level of hierarchy via concatenation. In this work, 7 layers for both the encoder and the decoder are used. Finally, the output of the decoder is spectral features $\mathbf{S}\in\mathbb{R}^{256\times256\times K}$ with the same size as the input spectrogram and $K$ channels.

\subsection{\label{subsec:endmodule}Fusion module}

Considering the visual feature from the vision network $\mathbf{v}\in[0,1]^{1\times K}$ and the spectral features from the audio network $\mathbf{S}\in\mathbb{R}^{256\times256\times K}$, the predicted spectrogram mask $\mathbf{M}^\prime\in[0,1]^{256\times256}$ is computed as:  

\begin{equation}
\mathbf{M}^\prime = \sigma\Big(\sum_{k=1}^K \alpha_{k} \textrm{v}_{k} \mathbf{S}_{k} + \beta\Big)
\label{eq:fusion}
\end{equation}
where $\mathbf{\alpha}\in\mathbb{R}^{1\times K}$ and $\beta\in\mathbb{R}$ are learnable parameters. $\sigma$ corresponds to the sigmoid activation function.

In this work, the multimodal fusion is performed at the end of the model as a linear combination like in PixelPlayer\cite{zhao2018sound}. Other fusion techniques exist to condition an audio network using visual features\cite{slizovskaia2020conditioned}. In early experiments, we inserted visual features in the middle of the audio network and in all decoder layers using a feature-wise linear modulation approach\cite{perez2018film}. However, model convergence turned out to be slower with no increase in performance.

\subsection{Learning Objective}

During the training of the network, the parameters of the model are optimized to reduce the binary cross entropy (BCE) loss between each value of the predicted frequency mask $\mathbf{M}^\prime$ and the ideal binary mask $\mathbf{M}^{\textrm{IBM}}$. Considering a mixture created by N sources, we first define the ideal binary mask $\mathbf{M}^{\textrm{IBM}}$ of the $i$-source as:

\begin{equation}
\mathbf{M}^{\textrm{IBM}}(t,f)=
\left\{
	\begin{array}{ll}
		1  & |S_{i}(t,f)| \geq |S_{n}(t,f)| \forall n\in \{1, \ldots, N\} \\
		0 & \mbox{otherwise},
	\end{array}
\right.
\end{equation}
where $S_{i}(t,f)=\textrm{STFT}(s_{i}(t))$. Then the loss function takes the form:

\begin{equation}
\mathcal{L} = \textrm{BCE}(\mathbf{M}^\prime, \mathbf{M}^{\textrm{IBM}})
\end{equation}

\subsection{\label{subsec:datasets}Dataset}

While there are lots of audio datasets, there is data scarcity of 3D videos of performing musicians. For the purposes of this work, we capture 3D videos of twelve performers playing different instruments: cello, double bass, guitar, saxophone, and violin. In addition, we separately collect audio recordings of these instruments from existing audio datasets\cite{montesinos2020solos, zhao2018sound}. Lastly, once all 3D video and audio data was collected, we randomly associate audio and 3D video for each instrument. This will enable the model to learn and perform source separation with unsynchronized 3D video and audio.

\textbf{3D video}: We capture a dataset consisting on 3D videos from several musicians playing different instruments: cello, doublebass, guitar, saxophone, and violin. Recordings were conducted using a single Azure Kinect DK placed one meter above the floor and capturing a frontal view of the musician at a distance of two meters. Azure Kinect DK comprises a depth camera and a color camera. The depth camera was capturing a $75\degree \times 65\degree$ field of view with a $640\times 576$ resolution while the color camera was capturing with a $1920\times 1080$ resolution. Both cameras were recording at 15 fps and Open3D library was then used to align depth and color streams and generate a point cloud for each frame. The full 3D video recordings span 1 hour of duration with an average of 12 different performers for each instrument.

We increase our 3D video dataset collecting 3D videos from \textit{small ensemble 3D-video database} \cite{thery2019anechoic} and \textit{Panoptic Studio} \cite{joo2017panoptic}. In \textit{small ensemble 3D-video database}, recordings are carried using three RGB-Depth Kinect v2 sensors. LiveScan3D \cite{kowalski2015livescan3d} and OpenCV libraries are then used to align and generate point clouds for each frame given each camera point of view and sensor data. The average video recordings is 5 minutes per instrument and a single performer per instrument. In \textit{Panoptic Studio}, recordings are carried out using ten Kinect sensors. In this case, recordings span two instrument categories: cello and guitar. The average video recordings is 2 minutes per instrument and a single performer per instrument.

We use 75\% of all 3D video recordings for training, 15\% for validation, and the remaining 10\% for testing. Note that the training, validation and testing sets are completely independent as there are no overlapping identities between them.

\textbf{Audio}: We collect audio recordings from \textit{Solos} \cite{montesinos2020solos} and \textit{Music} \cite{zhao2018sound} datasets to create an audio dataset that comprises audio sources from the aforementioned five different instruments. Both \textit{Solos} and \textit{Music} datasets gather audio from Youtube which provides recordings in a variety of acoustic conditions. In total, audio recordings span 30 hours of duration with an average amount of 72 recordings per instrument and a mean duration of 5 minutes per recording.

We use 75\% of all audio recordings for training, 15\% for validation, and the remaining 10\% for testing.

\subsection{\label{subsec:dataproc}Data preprocessing and data augmentation}

\subsubsection{3D videos}

\begin{figure}[!t]
	\centering
	\figline{\fig{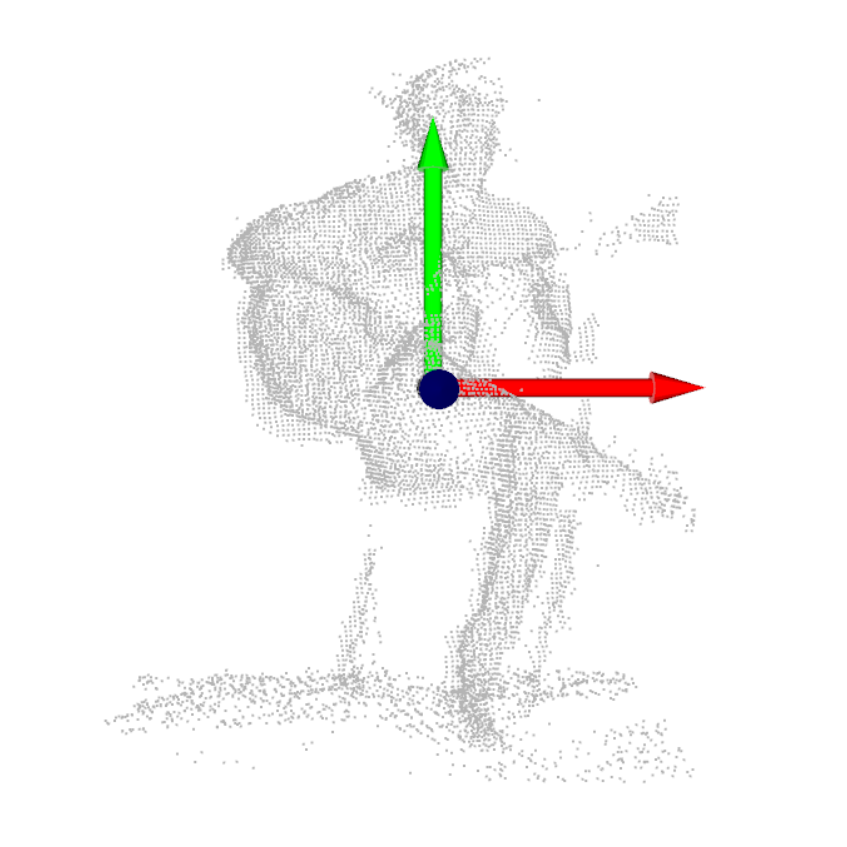}{0.5\linewidth}{(a)}}
	\figline{
	\leftfig{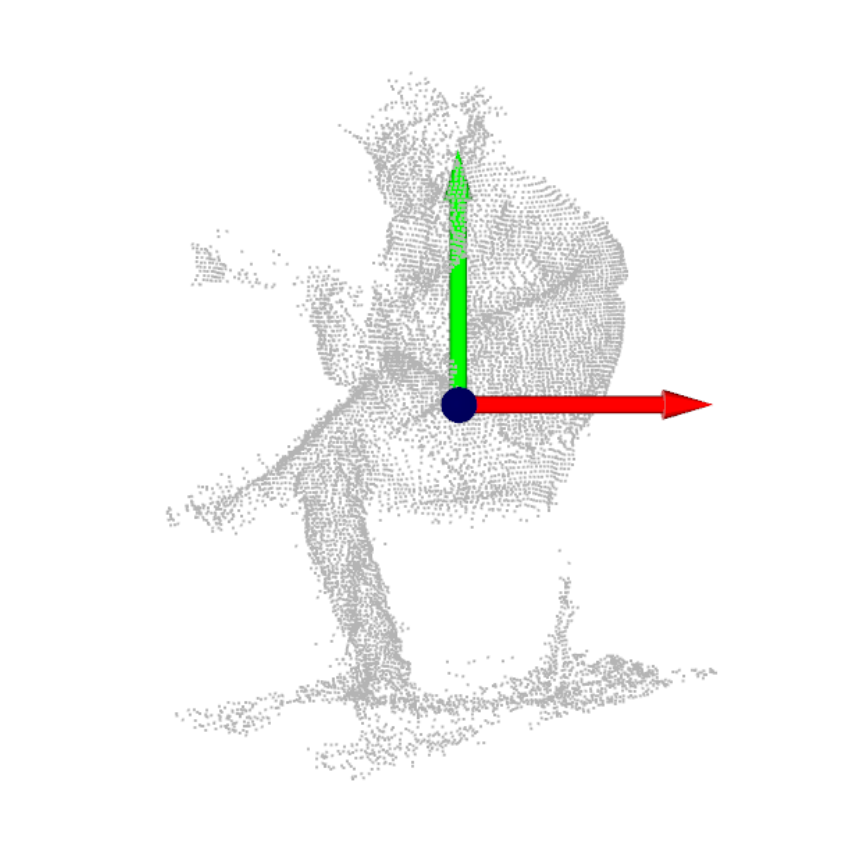}{0.5\linewidth}{(b)}
	\rightfig{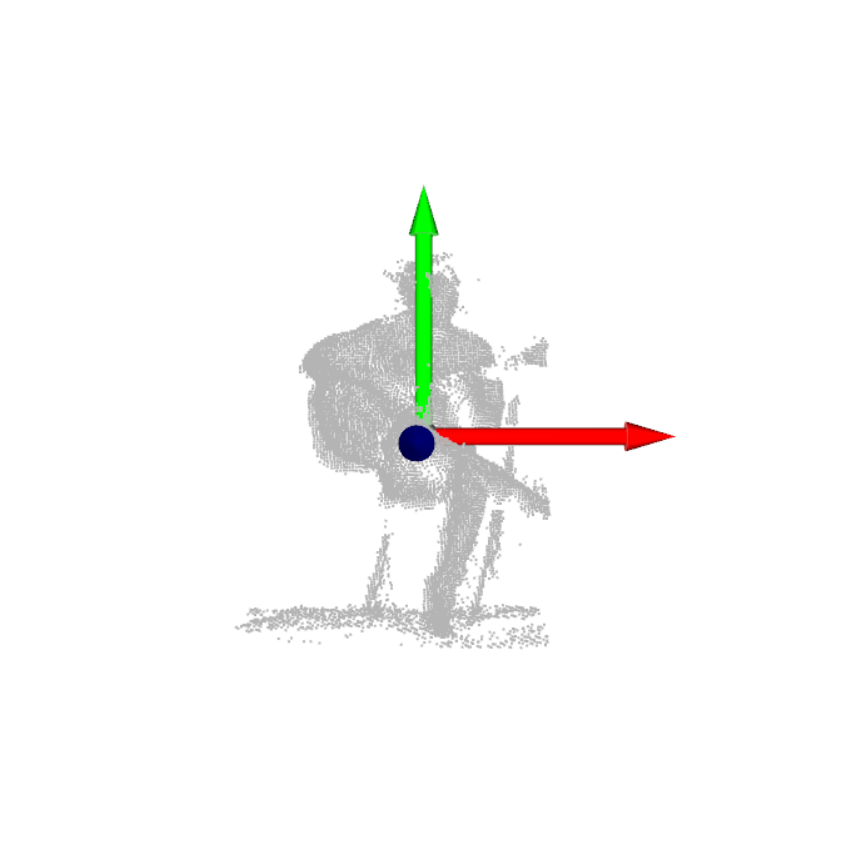}{0.5\linewidth}{(c)}}
	\figline{
	\leftfig{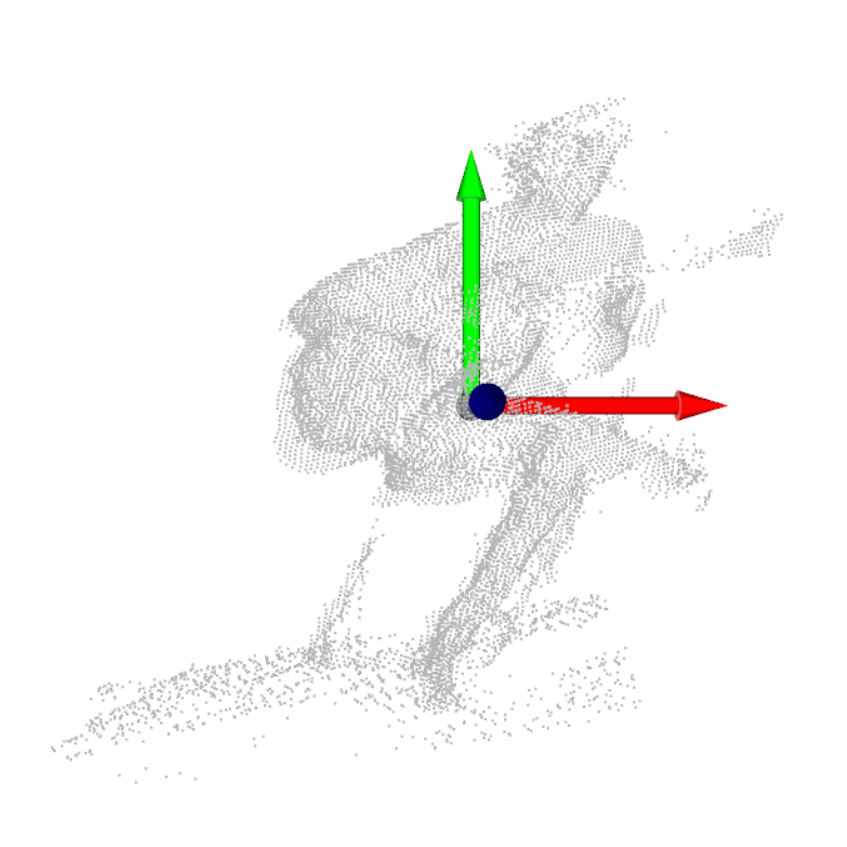}{0.5\linewidth}{(d)}
	\rightfig{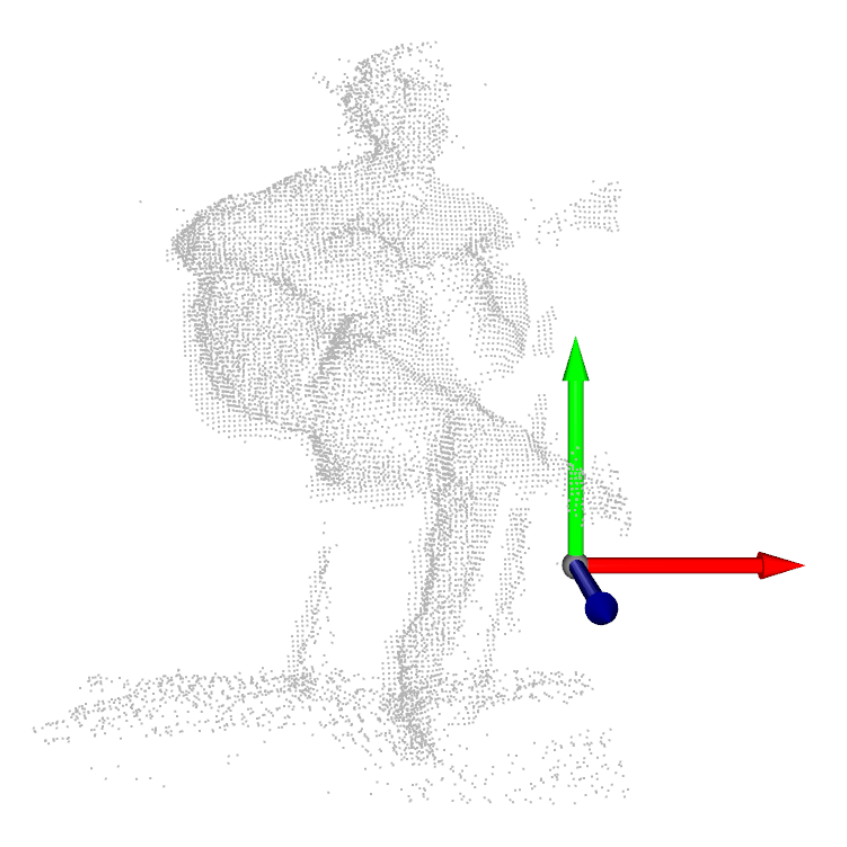}{0.5\linewidth}{(e)}}
	\caption{Illustration of the coordinate augmentations applied individually: (a) Original. (b) Random rotation. (c) Random scale. (d) Random shear. (e) Random translation. Red, green, and blue arrows correspond to x,y,z-axis respectively.
}
	\label{fig:points_transformation}
\end{figure}

The following preprocessing procedure is applied to all 3D videos. First, each frame, which consists of a musician point cloud, is scaled to fit within the cube $[-1, 1]^3$ and centered in the origin. Then, the axes representing the point clouds are set to have the same meaning for all the collected 3D videos. This is: the body face direction is z-axis, stature direction is y-axis, and the side direction is x-axis (see Fig.~\ref{fig:points_transformation}). 

During the training of the model, $F$ point cloud frames are randomly selected from each 3D video and a combination of augmentation operations are performed in both coordinates and color to increase the data diversity. After augmentation, resulting frames are fed into the vision network. 

We randomly rotate, scale, shear and translate the coordinates of the selected frames. We perform two rotations. First rotation is on the y-axis with a random angle ranging from $-\pi$ to $\pi$ and second rotation is on a random axis with a random angle ranging from $-\pi/6$ to $\pi/6$. Both axis and angles are sampled uniformly. The scaling factor is also uniformly sampled and ranges from 0.5 to 1.5 while the translation offset vector is sampled from a Normal distribution $\mathcal{N}_{1 \times3}(0, 0.4^2)$. Lastly, shearing is applied along all axes and the shear elements are sampled from $\mathcal{N}(0, 0.1^2)$. Figure~\ref{fig:points_transformation} illustrates the different augmentation operations applied to coordinates.

\begin{figure}[!t]
	\centering
	\figline{
	\leftfig{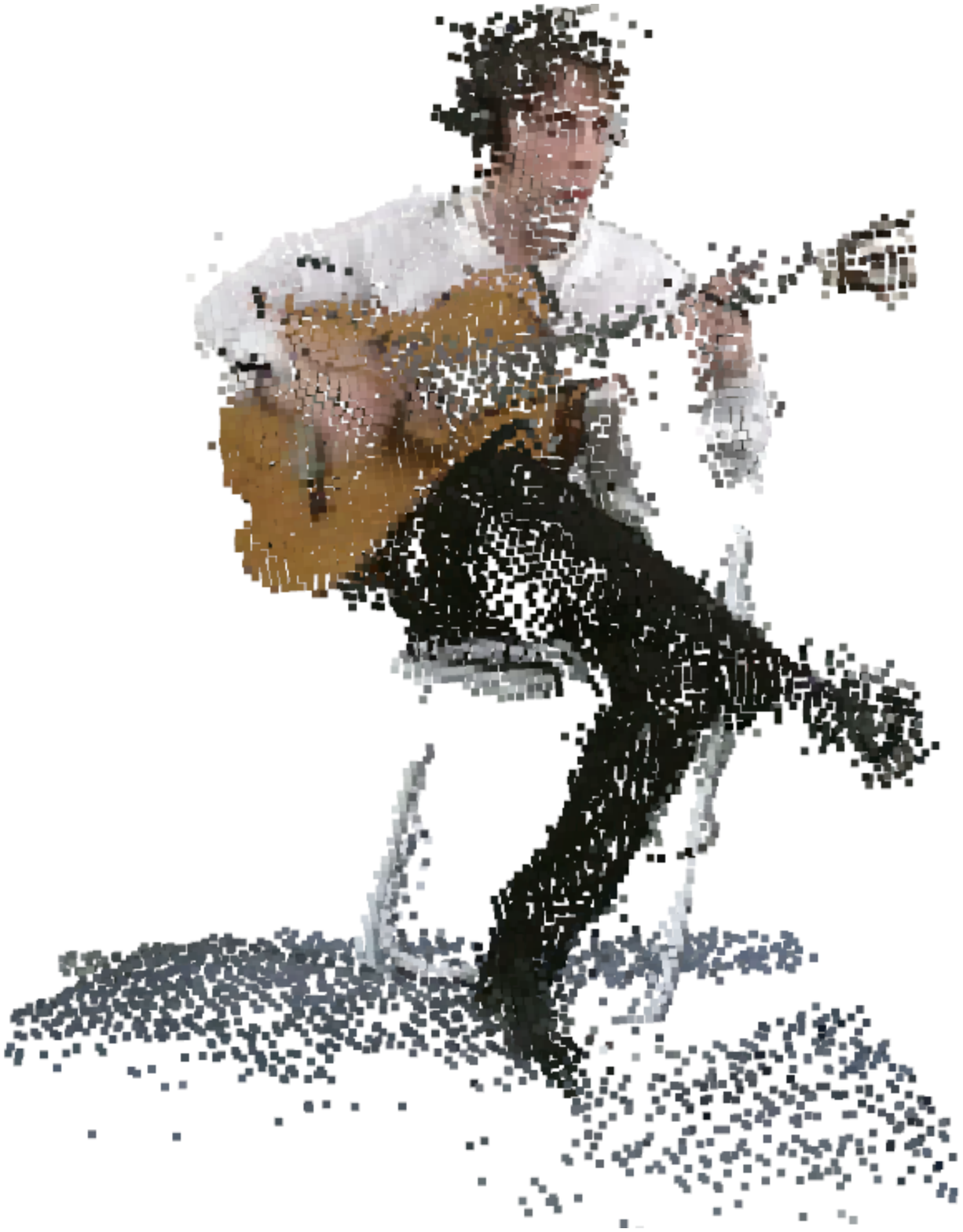}{0.4\linewidth}{(a)}
	\rightfig{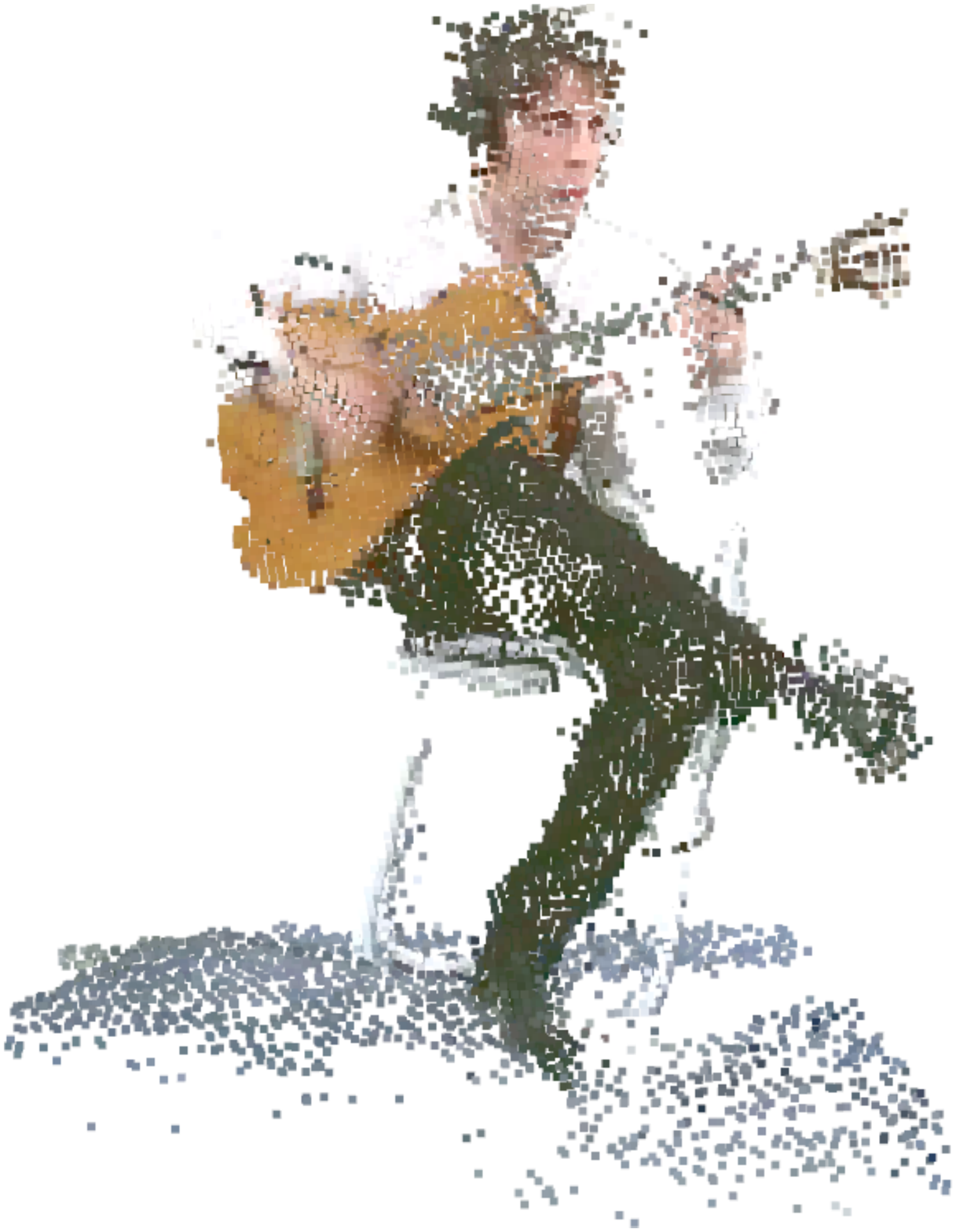}{0.4\linewidth}{(b)}}
	\figline{
	\leftfig{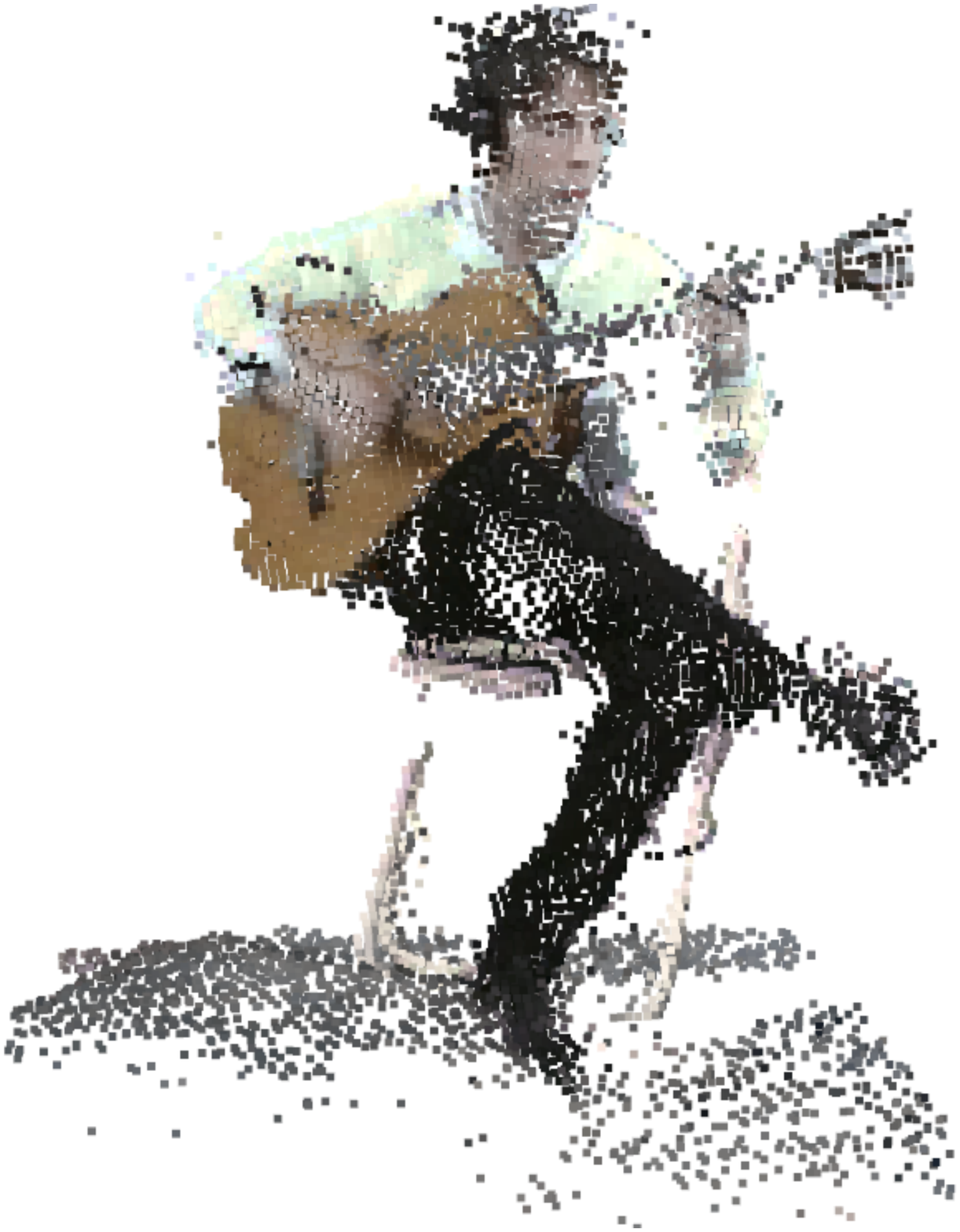}{0.4\linewidth}{(c)}
	\rightfig{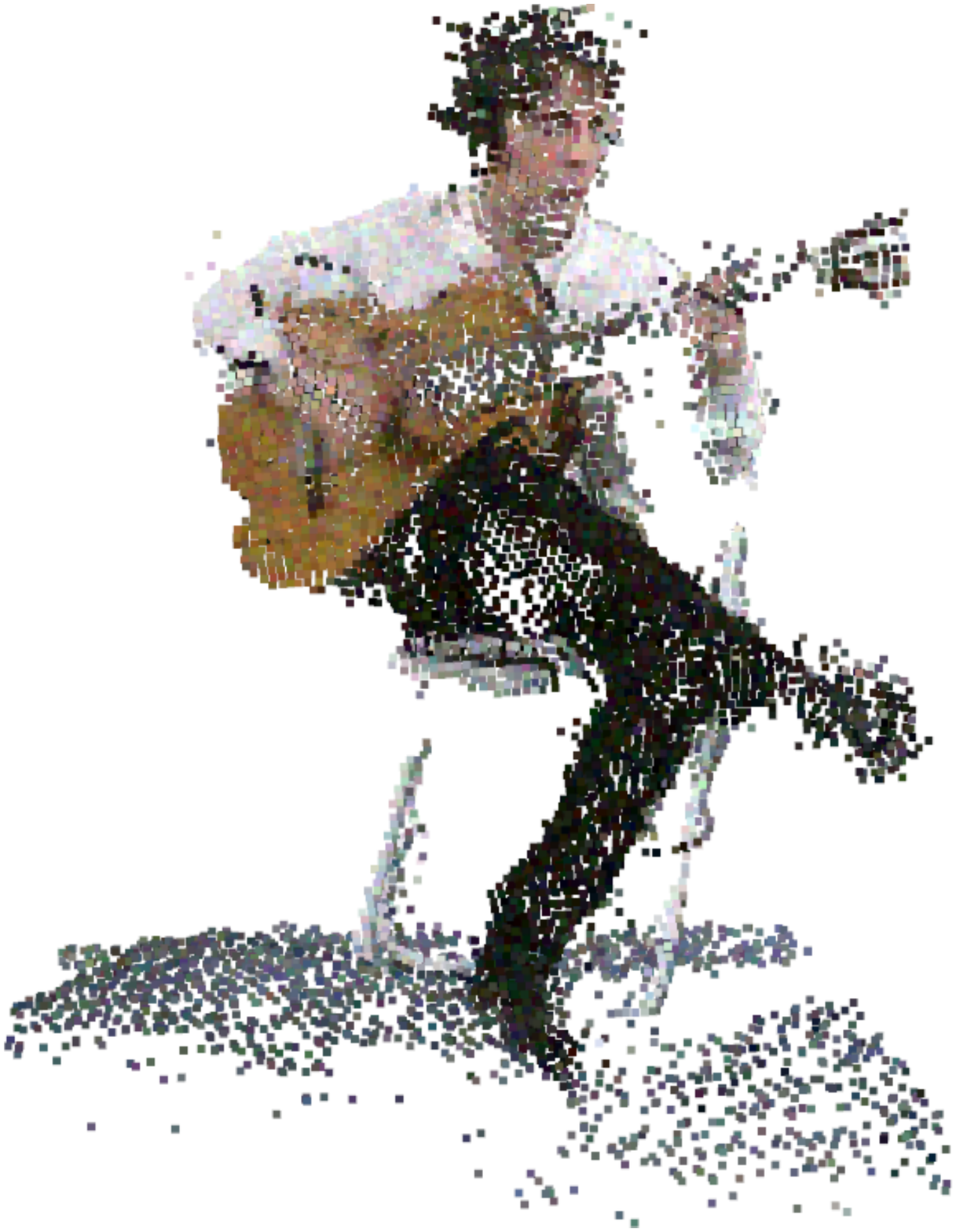}{0.4\linewidth}{(d)}}
	\caption{Illustration of the color augmentations applied individually: (a) Original. (b) Random value. (c) Random saturation. (d) Gaussian noise.}
	\label{fig:color_transformation}
\end{figure}

Regarding color, we distort the brightness and intensity of the selected frames. Specifically, we alter color value and color saturation with random amounts uniformly sampled ranging from -0.2 to 0.2 and -0.15 to 0.15 respectively. We also apply color distortion to each point via adding Gaussian noise on each rgb color channel. Gaussian noise is sampled from $\mathcal{N}(0, 0.05^2)$. Figure~\ref{fig:color_transformation} illustrates the different augmentation operations applied to colors.

After data augmentation, each frame is represented as a sparse tensor by discretizing the point cloud coordinates using a voxel size of 0.02. 
Performance of the model is further analyzed considering two different conditioning scenarios: when 3D videos consist only in a sequence of depth point cloud frames, or when 3D videos consist in a sequence of rgb-depth point cloud frames. In the case only depth information is available, we use the non-discretized coordinates as the feature vectors associated to each point. In the case rgb-depth information is available, we use the rgb values as the feature vectors associated to each point.

\subsubsection{Audio}

Audio recordings are converted to mono and resampled to 11025 Hz. During the training of the model, 6 seconds long snippets are randomly selected from each audio recording. Short-time Fourier transform is then applied using a Hann window with size 1022 and a hop size of 256. The magnitude of the time-frequency representation is selected and a log-frequency scale is applied. This results in a time frequency representation of size $256\times 256$.
As augmentation, audio snippets are scaled with respect to the amplitude by a random factor uniformly sampled ranging from $0.5$ to $1.5$.

\section{\label{sec:Results}Results}

\subsection{Implementation details}

Initially, we pretrain the vision network on the 3D object classification task modelnet40\cite{wu20153d} to assist the future learning process. As modelnet40 data consist on CAD models, point clouds are sampled from mesh surfaces of the object shapes. For the pretraining, we also discretize the coordinates setting the voxel size to 0.02 and the non-discretized coordinates are used as feature vectors associated to each point.

We train the whole model for 120k iterations using stochastic gradient descent with a momentum 0.9. We set $K=16$ as the number of channels extracted by both the vision and the audio network and we use a batch size of 40 samples. Learning rate is set to $1\times 10^{-4}$ for the vision network and to $1\times 10^{-3}$ for all the other learnable parameters of the model. We select the weights with less validation loss after the training process. Both training and testing are conducted on a single Titan RTX GPU. We use the Minkowski Engine\cite{choy20194d} for the sparse tensor operations within the vision network and PyTorch\cite{paszke2019pytorch} for the other operations required.

\subsection{Evaluation metrics}

We measure the audio separation performance of the model using the metrics from the \texttt{BSS\_eval} toolbox\cite{vincent2006performance}. In a first step, the toolbox decomposes the estimated signal \(\hat{s_{i}}\) as follows:
\begin{equation} 
\label{eq:estimated_source_decomposition}
\hat{s_{i}} = s_\text{target} + e_\text{interf} + e_\text{noise} + e_\text{artif}
\end{equation}
Where \(s_\text{target} = f(s_{i})\) is a version of the reference signal \({s_{i}}\) modified by an allowed distortion $f$. \(e_\text{interf}\) stands for the  interference coming from unwanted sources found in the original mixture, \(e_\text{noise}\) denotes the sensor noise, and \(e_\text{artif}\) refers to the burbling artifacts which are self-generated by the separation algorithm. Then, the objective metrics are computed as energy ratios in decibels (dB), where higher values are considered to be better. Source to distortion ratio (SDR) is defined as:
\begin{equation} 
\label{eq:SDR}
\textrm{SDR} = 10\log_{10} \frac{\lVert s_\text{target} \rVert^2}{\lVert e_\text{interf} + e_\text{noise} + e_\text{artif}\rVert^2}
\end{equation}
Source to interference ratio (SIR) is defined as:
\begin{equation} 
\label{eq:SIR}
\textrm{SIR} = 10\log_{10} \frac{\lVert s_\text{target} \rVert^2}{\lVert e_\text{interf}\rVert^2}
\end{equation}
Source to artifacts ratio (SAR) is defined as:
\begin{equation} 
\label{eq:SAR}
\textrm{SAR} = 10\log_{10} \frac{\lVert s_\text{target} + e_\text{interf} + e_\text{noise} \rVert^2}{\lVert e_\text{artif} \rVert^2}
\end{equation}

We also assess the separation performance using the scale-invariant source to distortion ratio\cite{le2019sdr} (SI-SDR). SI-SDR does not modify the reference signal \({s_{i}}\) and evaluates the separation performance within an arbitrary scale factor. SI-SDR between a signal $s_i$ and its estimated $\hat{s}_i$ is defined as: 
\begin{equation}
\textrm{SI-SDR} = 10\log_{10}\left(\frac{\|\alpha s_i\|^2}{\|\alpha s_i - \hat{s}_i\|^2}\right)
\end{equation}
where $\alpha = \textrm{argmin}_{\alpha}\|\alpha s_i - \hat{s}_i\|^2 = \hat{s}_i^Ts_i/\|s_i\|^2$.

\subsection{Baselines}

For comparison, we provide the separation performance of the model when conditioned on instrument labels. To this end, we change the visual feature vector $\mathbf{v}\in[0,1]^{1\times K}$ for a one-hot vector $\mathbf{h}\in[0,1]^{1\times 5}$ indicating the instrument we aim to separate. We set the number of spectral features extracted by the audio network accordingly, i.e.\ $K=5$. We further refer to this as label conditioning. Note that when using label conditioning, the one-hot vector will select one of the spectral features extracted by the audio network (see Eq.~\ref{eq:fusion}). Thus, label conditioning forces each spectral feature to correspond to a predicted spectrogram mask for each instrument. For that reason, label conditioning can be interpreted as assessing the modeling capabilities of the audio network alone for the separation task.

We also provide the upper bound separation performance when signals are estimated using the ideal binary masks (IBM). 

\subsection{Varying number of frames}

We are interested in evaluating the source separation performance depending on the number of 3D video frames $F$ used to condition the audio separation. We fix $N=2$ as the number of sources present in the mix and assess the separation performance when the vision network is provided with $F=1, 3, 5$ frames. We take consecutive frames with a distance of 1 second between them. In this case, performance is evaluated considering 3D videos that consist only in a sequence of depth frames, i.e. without color information. We use the non-discretized coordinates as the feature vectors associated to each point.

\begin{table}[ht]
\begin{ruledtabular}
\caption{Source separation performance for different number of frames.}

\begin{tabular}{cc|cccc}
 \# Frames & Method & SDR & SIR & SAR & SI-SDR\\
\hline
& label & 6.10 & 12.44 & 9.73 & 3.94 \\ \hline
1 & depth & 5.34 & 11.06 & 10.10 & 3.15 \\
3 & depth & 2.84 & 7.72 & 9.35 & 0.27  \\
5 & depth & 0.84 & 5.16 & 9.25 & -2.49  \\ \hline
& IBM & 15.20 & 21.63 & 16.84 & 14.47 \\

\end{tabular}
\label{table:frame_results}
\end{ruledtabular}
\end{table}

Results in Table \ref{table:frame_results} show an improved separation performance as the number of 3D frames provided to the vision network decreases. Best results are achieved when the separation is conditioned with a single 3D frame. This means that the vision network has difficulties to learn from multiple frames and cannot extract meaningful motion information from consecutive frames. Similar cases are found in 2D visual source separation studies where single frame conditioning provides strong separation performance\cite{slizovskaia2020conditioned, zhu2020separating}. 

Conditioning the model using a single 3D depth frame shows competitive performance compared with label conditioning. This suggest the model is able to recognize the musical instrument from the 3D depth frame as in the case where the instrument label is explicitly given as a one-hot vector.

\subsection{Varying number of sources}

\begin{figure*}[!t]
	\centering
	\includegraphics[width=\linewidth]{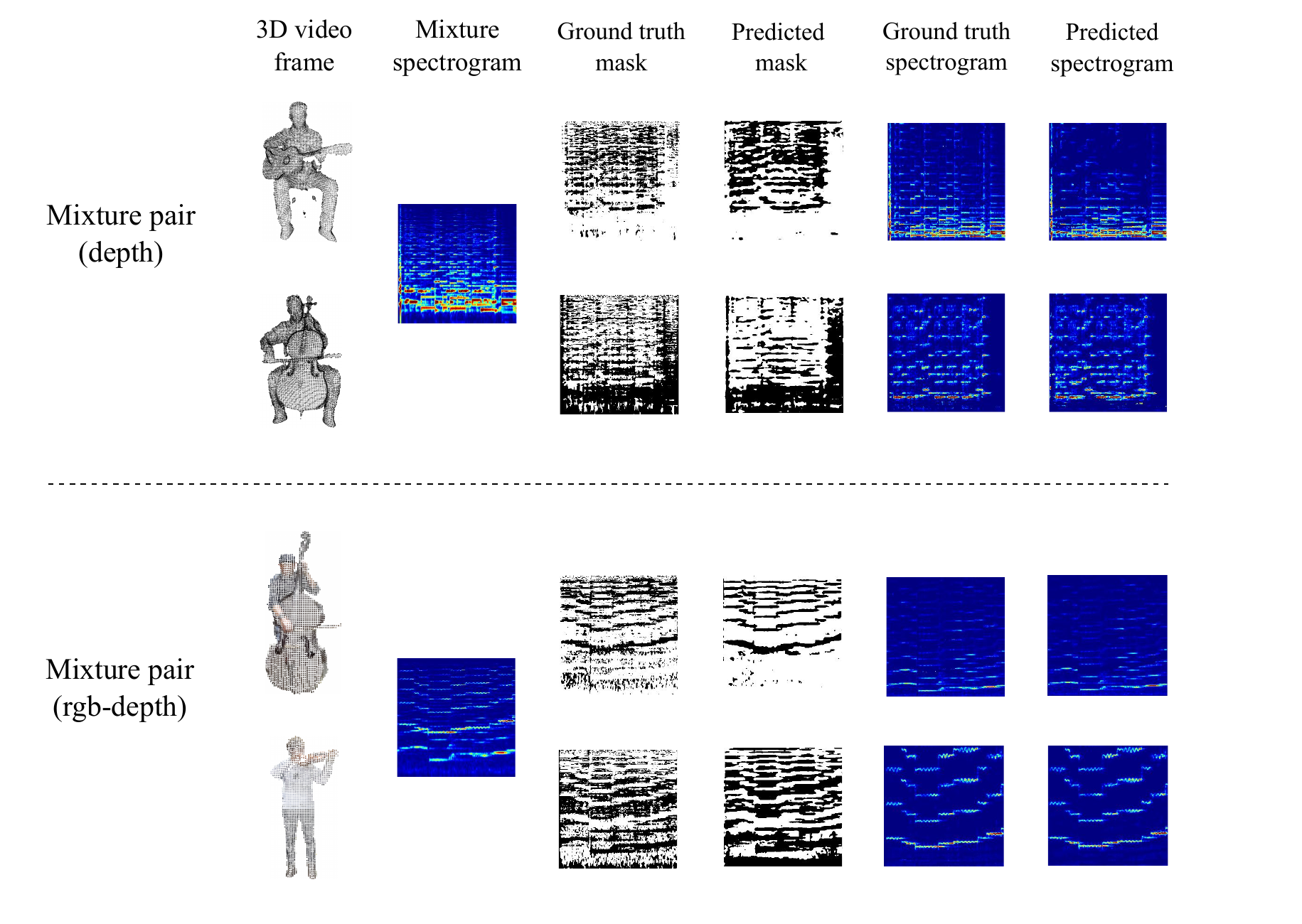}
	\caption{Visualization of the model separation performance using a single 3D video frame and two sources in the mix.}
	\label{fig:visualization_predictions}
\end{figure*}

We evaluate the source separation performance with regard to the number of sources present in the mix. Considering best results from previous section, we fix $F=1$ as the number of frames provided to the vision network and asses the separation performance considering $N=2, 3, 4$ sources in the mix. In this case, performance is evaluated considering two types of 3D videos: when 3D videos consist only in a sequence of depth frames, and when 3D videos consist in a sequence of rgb-depth frames. In the case only depth information is available, we use the non-discretized coordinates as the feature vectors associated to each point. In the case rgb-depth information is available, we use the rgb values as the feature vectors associated to each point.

\begin{table}[ht]
\caption{\label{tab:frames_results} Source separation performance for different number of sources (musical instruments) in the audio mixture.}

\begin{ruledtabular}
\begin{tabular}{cc|cccc}
 \# Sources & Method & SDR & SIR & SAR & SI-SDR \\
 \hline
 \multirow{4}{*}{2} & label & 6.10 & 12.44 & 9.73 & 3.94 \\
 & depth & 5.34 & 11.06 & 10.10 & 3.15 \\
 & rgb-depth & 6.14 & 12.29 & 9.97 & 3.98 \\
 & IBM & 15.20 & 21.63 & 16.84 & 14.47 \\
 \hline
 \multirow{4}{*}{3} & label & 1.35 & 8.10 & 5.64 & -2.75 \\
 & depth & -0.06 & 6.63 & 5.17 & -5.01 \\
 & rgb-depth & 0.86 & 7.27 & 5.69 & -3.39 \\
 & IBM & 11.80 & 19.55 & 12.93 & 10.56 \\
 \hline
 \multirow{4}{*}{4} & label & -1.76 & 4.21 & 3.72 & -7.03 \\
 & depth & -2.48 & 3.34 & 4.05 & -8.26 \\
 & rgb-depth & -3.39 & 2.79 & 3.48 & -9.78 \\
 & IBM & 9.14 & 16.65 & 10.26 & 7.66 \\
\end{tabular}
\label{table:source_results}
\end{ruledtabular}
\end{table}

Regarding 3D visual conditioning, results reported in Table \ref{table:source_results} show a better separation performance when the model is conditioned using rgb-depth information for $N=2, 3$ sources in the mix. Interestingly, when $N=4$ better separation is achieved using only depth information.

3D visual conditioning provides similar separation performance as label conditioning when $N=2$ sources in the mix. Specifically, 3D rgb-depth conditioning slightly outperforms label conditioning. This is likely due to the linear combination within the fusion module which provides more expressivity to the model when combining both vision and audio features.

As we increase the number of sources in the mix, i.e.\ $N=3, 4$, label conditioning performs better than 3D visual conditioning and its difference in performance increases with the number of sources. This shows that as more sources are present in the mix, it becomes more difficult for the network to jointly learn from 3D representations and the audio information in order to perform source separation. Figure~\ref{fig:visualization_predictions} depicts qualitative results of the proposed model when two sources are present in the mix.

\section{\label{sec:Discussion}Discussion}

The work presented here indicates the effectiveness of a proposed learning model to perform music source separation conditioned on 3D visual information. Specifically, 3D visual conditioning achieves competitive audio separation performance compared to label conditioning. 

Best separation performance is achieved when conditioning the model using a single 3D frame. This makes sense in the current scenario where 3D video and audio data are artificially associated and no synchronization between both modalities is provided. In addition, this can be understood in relation to the literature where 2D single frame conditioning \cite{slizovskaia2020conditioned, zhu2020separating} already achieves competitive results given the difficulties of learning  temporal information from multiple frames when no motion information is given. However, higher performance is expected by exploiting motion information as other 2D audio-visual models do in audio-visual data synchronization scenarios\cite{zhao2019sound, gan2020music}. Moreover, adopting a curriculum learning strategy\cite{bengio2009curriculum} may be beneficial to improve the model separation performance when the number of sources present in the mix increases\cite{zhao2019sound, slizovskaia2020conditioned}. Regarding the type of 3D information used to condition the separation, rgb-depth appears to perform better overall although only depth is better in the case when four sources are present in the mix.

The presented approach investigates to what extent point clouds can be used for guiding source separation, thus enabling potential applicability in the VR/AR domain. VR/AR systems benefit from knowing the sound and the location of each source in a 3D environment to simulate basic acoustic phenomena such as interaural time differences and interaural level differences. 
3D visual acquisition systems allow to capture the 3D environment, but the recorded sound sources are usually far from the acoustic sensor and therefore sound needs to be separated and associated to each source for further auralization. By learning jointly from 3D point clouds and audio, the proposed model is able to separate the sounds, given the sources of interest are represented as 3D point clouds. This opens a new research direction which can contribute to better simulation of acoustic phenomena in VR/AR applications.

\section{Conclusions}

In this paper, a deep learning model for music source separation conditioned on 3D point clouds has been proposed. The model jointly learns from 3D visual information and audio in a self-supervised fashion. Results show the effectiveness of the model when conditioned using a single frame and for a modest number of sources present in the audio mixture. 3D visual conditioning, as opposed to 2D, enables potential applicability in virtual and augmented reality applications where information about the sound and the location of each source in the 3D environment is required for further auralization. As future work, exploiting also spatial audio cues caused by the location of sound sources in the 3D environment may help to separate same instrument sources. In addition, multimodal models that jointly learn from audio and point clouds could improve auditory 3D scene understanding.

\begin{acknowledgments}
This project has received funding from the European Union's Horizon 2020 research and innovation programme under the Marie Sk\l{}odowska-Curie grant agreement No 812719. The authors thank A. Mayer for technical support.
\end{acknowledgments}

%  End of title page for Preprint option --------------------------------- %

%% See preprint.tex/.pdf or reprint.tex/.pdf for many examples

% -------------------------------------------------------------------------------------------------------------------
%   Appendix  (optional)

%\appendix
%\section{Appendix title}

%If only one appendix, please use
%\appendix*
%\section{Appendix title}

\bibliography{sampbib}

%=======================================================

%Use \bibliography{<name of your .bib file>}+
%to make your bibliography with BibTeX. 

%=======================================================

\end{document}